\documentclass[10pt,aps,prd,twocolumn,preprintnumbers,superscriptaddress,amsmath,amssymb,nofootinbib,floatfix]{revtex4-1}

\usepackage[hidelinks]{hyperref}
\usepackage{mathtools}
\usepackage{aas_macros}
\usepackage[capitalize]{cleveref}
\usepackage{amsmath}
\usepackage{amssymb}
\usepackage{amstext}
\usepackage{graphicx}
\usepackage{siunitx}

\Crefname{equation}{Equation}{Equations}
\crefname{equation}{Eq.}{Eqs.}
\Crefname{figure}{Figure}{Figures}
\crefname{figure}{Fig.}{Figs.}
\Crefname{table}{Table}{Tables}
\crefname{table}{Tab.}{Tabs.}
\Crefname{section}{Section}{Sections}
\crefname{section}{Sec.}{Secs.}

\newcommand{\Z}{\mathbb{Z}}
\newcommand{\R}{\mathbb{R}}
\newcommand{\p}{\mathrm{p}}
\newcommand{\kkd}{k_{-}}
\newcommand{\kusr}{k_{+}}
\newcommand{\etat}{\eta_\mathrm{t}}
\newcommand{\m}{m_\p}
\newcommand{\mm}{m_\p^2}
\newcommand{\prm}[1]{{#1}^{\prime}}
\newcommand{\dprm}[1]{{#1}^{\prime\prime}}
\newcommand{\sft}{\beta}
\renewcommand{\d}[2][]{\operatorname{d}^{#1}\!{#2}}

\newcommand{\Planck}{\textit{Planck}}
\newcommand{\conformalH}{\mathcal{H}}
\newcommand{\deriv}{\prm}
\newcommand{\dderiv}{\dprm}

\usepackage{xcolor}

\hypersetup{colorlinks, linkcolor={blue}, citecolor={blue}, urlcolor={blue}}
\usepackage[font=small, labelfont=bf, justification = raggedright]{caption}
\usepackage{graphicx}

\usepackage{subcaption}
\usepackage{bm}

\begin{document}

\preprint{Prepared for submission to Phys. Rev. D}
\title{Analytical approximations for curved primordial power spectra}

\author{Ayngaran Thavanesan}
\email[]{at735@cantab.ac.uk}
\affiliation{Astrophysics Group, Cavendish Laboratory, J.J.Thomson Avenue, Cambridge, CB3 0HE, United Kingdom}
\affiliation{Kavli Institute for Cosmology, Madingley Road, Cambridge, CB3 0HA, United Kingdom}

\author{Denis Werth}
\email[]{denis.werth@ens-paris-saclay.fr}
\affiliation{Astrophysics Group, Cavendish Laboratory, J.J.Thomson Avenue, Cambridge, CB3 0HE, United Kingdom}
\affiliation{Kavli Institute for Cosmology, Madingley Road, Cambridge, CB3 0HA, United Kingdom}
\affiliation{\'{E}cole Normale Sup\'{e}rieure Paris-Saclay, Avenue des Sciences, Gif-sur-Yvette, 91190, France}

\author{Will Handley}
\email[]{wh260@mrao.cam.ac.uk}
\affiliation{Astrophysics Group, Cavendish Laboratory, J.J.Thomson Avenue, Cambridge, CB3 0HE, United Kingdom}
\affiliation{Kavli Institute for Cosmology, Madingley Road, Cambridge, CB3 0HA, United Kingdom}
\affiliation{Gonville \& Caius College, Trinity Street, Cambridge, CB2 1TA, United Kingdom}

\date{\today}

\begin{abstract}
    \vspace{5pt}
    We extend the work of \citet{Contaldi} and derive analytical approximations for primordial power spectra arising from models of inflation which include primordial spatial curvature. These analytical templates are independent of any specific inflationary potential and therefore illustrate and provide insight into the generic effects and predictions of primordial curvature, manifesting as cut-offs and oscillations at low multipoles and agreeing with numerical calculations. We identify through our analytical approximation that the effects of curvature can be mathematically attributed to shifts in the wavevectors participating dynamically.
    \vspace{20pt}
\end{abstract}

\maketitle

\section{Introduction}\label{sec:introduction}
The inflationary scenario~\citep{Starobinskii1979, Guth1981, Linde1982} was invoked to resolve several issues within the basic Hot Big Bang model, and it is upon this that the current concordance cosmology, Lambda cold dark matter ($\Lambda$CDM), is built. Through a brief period of rapid expansion at early times, the inflationary framework successfully predicts the minimal present-day curvature, as well as the generation and growth of nearly scale-invariant adiabatic scalar perturbations. These perturbations then manifest themselves in the cosmic microwave background (CMB), as anisotropies, giving us the measured spectrum we observe on the sky today~\citep{planck_parameters, planck_inflation2018}.

If one is to study inflation in a theoretically complete manner, one cannot assume that the Universe was flat at the start of the expansion.  Furthermore, the presence of small discrepancies at low multipoles in the spectrum of the CMB~\citep{planck_isotropy} from those predicted by flat inflationary dynamics, motivate a study of the effects of primordial curvature. Typically the spectra contain generic cut-offs and oscillations within the observable window for the level of curvature allowed by current CMB measurements. Previous numerical calculations~\citep{Handley_2019} of such models have shown that the primordial power spectra generated for curved inflating universes provide a better fit to current data. A Bayesian discussion as to the level of fine-tuning in these curved inflationary models (or lack thereof) can be found in detail in~\citep{Hergt2, Lukas_2020}.

Additionally, the introduction of a small amount of late-time curvature, creating a $K\Lambda$CDM cosmology~\citep{Ellis_2003, lasenbyclosed}, has been suggested as a potential resolution to the tensions observed between datasets probing the early Universe and those that measure late-time properties~\citep{Riess2018, Joudaki2016, Kohlinger2017, Hildebrandt2016, DESParameters2017, tension, Philcox2020, H0T0tension}. Planck 2018 data without the lensing likelihood~\citep{planck_lensing} presents relatively strong evidence for a closed Universe~\citep{planck_parameters}. Adding in lensing and Baryon acoustic oscillation data~\citep{SDSS, SDSS2, SDSS3} reduces this evidence considerably, but it remains an open question as to why the CMB alone prefers universes with positive spatial curvature. Whilst interpretations of the level of experimental support for a moderately curved present-day Universe differ~\citep{2019arXiv190809139H, 2020NatAs...4..196D, 2020arXiv200206892E}, Universe models with percent-level spatial curvature remain compatible with CMB datasets. The appearance of any present-day curvature is arguably incompatible with eternal inflation, and strongly constrains the total amount of inflation, providing a powerful justification for just-enough-inflation theories~\citep{kinetic_dominance, Hergt1, just_enough_inflation, just_enough_inflation2, BVS1, BVS2, Avis_2020}.

In this paper, we generalise the approach of \citet{Contaldi} to the curved case, obtaining analytical background solutions and primordial power spectra for universes including spatial curvature. The approximation models the background Universe as beginning in a kinetically dominated regime, followed by an instantaneous transition to a regime with no potential dependence, which we term \emph{ultra-slow-roll}. Despite such an idealised situation, this simple approximation qualitatively reproduces the exact spectrum obtained by numerical computation, with the notable advantage of using this method that the results are independent of the scalar field potential. The analytic solutions yield better insight into the physics and effects of curvature on the primordial Universe which may potentially be overlooked through a purely numerical approach. 

This paper is organised as follows. In \cref{sec:background} the conformal time equivalent of the background equations and general Mukhanov-Sasaki equation for curved inflating universes are presented. We solve the curved Mukhanov-Sasaki equation and plot the corresponding spectra for our potential-independent curved inflationary model in \cref{sec:computingCurvedSpectra}. This is proceeded by a discussion of our results in \cref{sec:discussion}, after which we present our conclusions in \cref{sec:conclusions}. Supplementary material, such as PYTHON code for generating figures and Mathematica scripts for computer algebra, is found at~\citep{Zenodo}.

\vfill
\pagebreak
\section{Background}\label{sec:background}
In this section we establish notation and sketch a derivation of the background and first-order perturbation equations in conformal time. Further detail and explanation may be found in~\citep{1992PhR...215..203M, Lesgourgues:2013bra, Baumann}.

The action for a single-component scalar field minimally coupled to a curved spacetime is
\begin{equation}
    S = \int \d[4]{x}\sqrt{|g|}\left\{ \frac{1}{2}R + \frac{1}{2}\nabla^\mu\phi\nabla_\mu\phi - V(\phi)\right\}.
    \label{eqn:action}
\end{equation}

Extremising this action generates the Einstein field equations and a conserved stress energy tensor. 

In accordance with the cosmological principle, the solutions to the Einstein field equations are assumed to be homogeneous and isotropic at zeroth order. One then perturbatively expands about the homogeneous solutions to first order in the Newtonian gauge. 

In conformal time and spherical polar coordinates in the Newtonian gauge, the metric may be written as
\begin{gather}
    \d{s}^2 = {a(\eta)}^2[(1+2\Phi)\d{\eta}^2 - (1-2\Psi) (c_{ij}+h_{ij})\d{x}^i \d{x}^j],
    \nonumber\\
    c_{ij}\d{x^i}\d{x^j} = \frac{\d{r}^2}{1-Kr^2} + r^2(\d{\theta}^2  + \sin^2\theta\d{\phi}^2),
    \label{eqn:metric}
\end{gather}
where $K\in\{+1,0,-1\}$ denotes the sign of the curvature of the Universe: flat ($K=0)$, open ($K=-1$) and closed ($K=+1$).\footnote{Note this is opposite to the curvature density parameter $\Omega_K$, $K=+1\Rightarrow \Omega_K<0$} The longitudinal metric perturbation $\Phi$ and curvature metric perturbation $\Psi$ along with the perturbation to the field $\delta\phi$ are scalar perturbations, whilst $h_{ij}$ is a divergenceless, traceless tensor perturbation with two independent polarisation degrees of freedom. The covariant spatial derivative on comoving spatial slices is denoted with a Latin index as~$\nabla_i$.

By taking the ($00$)-component of the Einstein field equations and the ($0$)-component of the conservation of the stress-energy tensor, one can show that the background equations for a homogeneous Friedmann-Robertson-Walker (FRW) spacetime with material content defined by a scalar field are
\begin{align}
    \conformalH^2  + K &= \frac{1}{3\mm}\left( \frac{1}{2}{\deriv{\phi}}^2 + a^2V(\phi) \right),
    \label{eqn:friedmann}\\
    0 &= \dderiv{\phi} + 2 \conformalH\deriv{\phi} + a^2\frac{\d{}}{\d{\phi}}V(\phi),
    \label{eqn:klein_gordon}
\end{align}
where $\conformalH=\deriv{a}/a$ is the conformal Hubble parameter, $\m$ is the Planck mass, $\phi$ is the homogeneous value of the scalar field, $V(\phi)$ is the scalar potential, $a$ is the scale factor and primes indicate derivatives with respect to conformal time $\eta$ defined by $\mathrm{d}\eta = \mathrm{d}t/a$. A further useful relation to supplement \cref{eqn:friedmann,eqn:klein_gordon} is
\begin{equation}
    \deriv{\conformalH} = -\frac{1}{3\mm}\left( {\deriv{\phi}}^2 - a^2V(\phi) \right),
    \label{eqn:raychaudhuri}\\
\end{equation}
which is derived from the trace of the Einstein field equations. For the remainder of this paper we set the Planck mass to unity ($\m=1$), but note that one may reintroduce $\m$ at any time by replacing $\phi\to\phi/\m$, $V\to V/\mm$.

Another useful physical perturbation to consider is the gauge-invariant comoving curvature perturbation
\begin{equation}
    \mathcal{R} = \Psi +\frac{\conformalH}{\deriv\phi}\delta\phi.
\end{equation}

The equation of motion for this quantity is termed the Mukhanov-Sasaki equation. To derive this equation for curved universes, one can take a direct perturbative approach as that introduced by \citet{1992PhR...215..203M}. This computation has been performed historically by~\citep{2003astro.ph.10127Z,Gratton_2002,Ratra_2017,Bonga_2016,Bonga_2017,Akama_2019,Ooba_2018}. One can also arrive at \cref{eqn:conformalcurvedMS} via the Mukhanov action, by following the notation of~\citet[Appendix B]{Baumann} and generalising the Arnowitt-Deser-Misner (ADM) formalism~\citep{Arnowitt_2008,Prokopec_2012} to the curved case.

Employing both approaches, a general version of the Mukhanov-Sasaki equation for curved universes was computed by \citet{Handley_2019} in cosmic time. Extending these calculations to conformal time, we now show that the curved Mukhanov-Sasaki equation is given by
\begin{gather}
    \left( \mathcal{D}^2 - K\mathcal{E} \right) \dderiv{\mathcal{R}}
    + \left( \left(\frac{{\deriv{\phi}}^2}{\conformalH} + \frac{2\dderiv{\phi}}{\deriv{\phi}} - \frac{2K}{\conformalH}\right)\mathcal{D}^2 - 2K \conformalH\mathcal{E} \right) \deriv{\mathcal{R}} \nonumber\\
    + \left(-\mathcal{D}^4 + K\left( \frac{2K}{\conformalH^2} - \mathcal{E} + 1 -\frac{2\dderiv{\phi}}{\deriv{\phi}\conformalH}  \right)\mathcal{D}^2 + K^2\mathcal{E}\right) \mathcal{R} =0,
    \nonumber\\
    \mathcal{D}^2 = \nabla_i\nabla^i + 3 K, \quad
    \mathcal{E} = \frac{{\deriv{\phi}}^2}{2\conformalH^2},
    \label{eqn:conformalcurvedMS}
\end{gather}
where primes denote derivatives with respect to conformal time. \cref{eqn:conformalcurvedMS} can be expressed in a more familiar form by Fourier decomposition and a redefinition of variables. In the flat case one normally redefines variables in terms of the Mukhanov variable $v = z\mathcal{R}$, where $z = a\deriv{\phi}/\conformalH$. In the curved case, this is impossible, but one can define a wavevector-dependent $\mathcal{Z}$ and $v$ via
\begin{gather}
    v = \mathcal{Z}\mathcal{R}, \quad \text{and} \quad \mathcal{Z} = \frac{a\deriv{\phi}}{\conformalH}\sqrt{\frac{\mathcal{D}^2}{\mathcal{D}^2 - K\mathcal{E}}}.
    \label{eqn:curvedMSdefinitions1}
\end{gather}

Fourier decomposition acts to replace the $\mathcal{D}^2$ operator in \cref{eqn:conformalcurvedMS} with its associated scalar wavevector expression~\citep{Lesgourgues:2013bra}
\begin{align}
    \mathcal{D}^2 &\leftrightarrow -\mathcal{K}^2(k)+3K, \\
    \mathcal{K}^2(k) &= 
    \left\{\begin{array}{llll}
        k^2, & k \in \R, & k > 0, & K = 0,-1, \\
        k(k+2), & k \in \Z, & k > 2, & K = +1. \\
    \end{array}\right.
    \label{eqn:wavevectors}
\end{align}

After some algebraic manipulation, the curved Mukhanov-Sasaki equation may be written as
\begin{equation}
    \dderiv{v}_k + \left[ \mathcal{K}^2 - \left(\frac{\dderiv{\mathcal{Z}}}{\mathcal{Z}} + 2K + \frac{2K\deriv{\mathcal{Z}}}{\conformalH\mathcal{Z}}\right) \right] v_k = 0. \\
    \label{eqn:GeneralCurvedMS}
\end{equation}

\pagebreak
\section{Analytical primordial power spectra for curved universes}\label{sec:computingCurvedSpectra}
To obtain spectra for curved inflating cosmologies, we will now generalise to the curved case an approximate analytical approach first applied by \citet{Contaldi}. For our model we assume a pre-inflationary kinetically dominated regime defined by ${\deriv{\phi}}^2 \gg a^2 V(\phi)$. We then invoke an instantaneous transition to a regime where the scalar field motion has significantly slowed ${\deriv{\phi}}^2 \ll a^2 V(\phi)$, and the standard slow-roll constraints to solve the horizon problem are satisfied. This rather brutal approximation has the advantage that it does not depend on a specific potential choice $V(\phi)$ and illustrates the effects of curvature on the primordial power spectrum. Furthermore, this model grants a framework within which potential dependence can be added via higher order terms in the solutions for curved inflationary dynamics.

In \cref{sec:Solving_Background} we provide analytic solutions and power series expansions for the background variables in the two regimes. In \cref{sec:Solving_CurvedMS_KD,sec:Solving_CurvedMS_USR} we derive analytic solutions for the mode equations in each regime, and match these together at the transition point.  \cref{sec:PowerSpectrum} then uses the freeze-out values of the ultra-slow-roll solution to produce our analytic template in \cref{eqn:PowerSpectrumR_final}.

To avoid confusion, note that we work in a convention where the conformal time $\eta=0$ is at the singularity, \textit{i.e.} $a(\eta = 0) = 0$, which is different from~\citet{Contaldi}, who place $\eta=0$ at the transition time. Also note that for our convention the scale factor $a$ has units of length; we work with a convention where scale factor $a \neq R/R_0$, and hence does not have the usual normalisation to unity at the present day, i.e.\ at redshift $z=0$, $a(z=0) \equiv a_0 \neq 1$. It has been shown by \citet{Agocs_2020} that through this redefinition of the present day scale factor, the comoving wavenumber and the physical scale of the curvature perturbation today differ by a factor of $a_0$.

\subsection{Background dynamics}\label{sec:Solving_Background}
To solve for the background variables $a$, $\mathcal{H}$ and $\phi$ we can rearrange \cref{eqn:friedmann,eqn:raychaudhuri} into two useful forms
\begin{align}
    \deriv{\conformalH} + 2\conformalH^2 + 2K &= a^2 V(\phi)\label{eqn:kd},\\
    \deriv{\conformalH} - \conformalH^2 - K &= -\frac{1}{2}{\deriv{\phi}}^2\label{eqn:usr}.
\end{align}

In the initial stages of kinetic dominance ${\deriv{\phi}}^2 \gg a^2 V(\phi)$, we can neglect the right-hand-side of \cref{eqn:kd}, and similarly in the ultra-slow-roll stage ${\deriv{\phi}}^2 \ll a^2 V(\phi)$ we can set the right hand side of \cref{eqn:usr} similarly to zero.

\begin{figure}
    \includegraphics{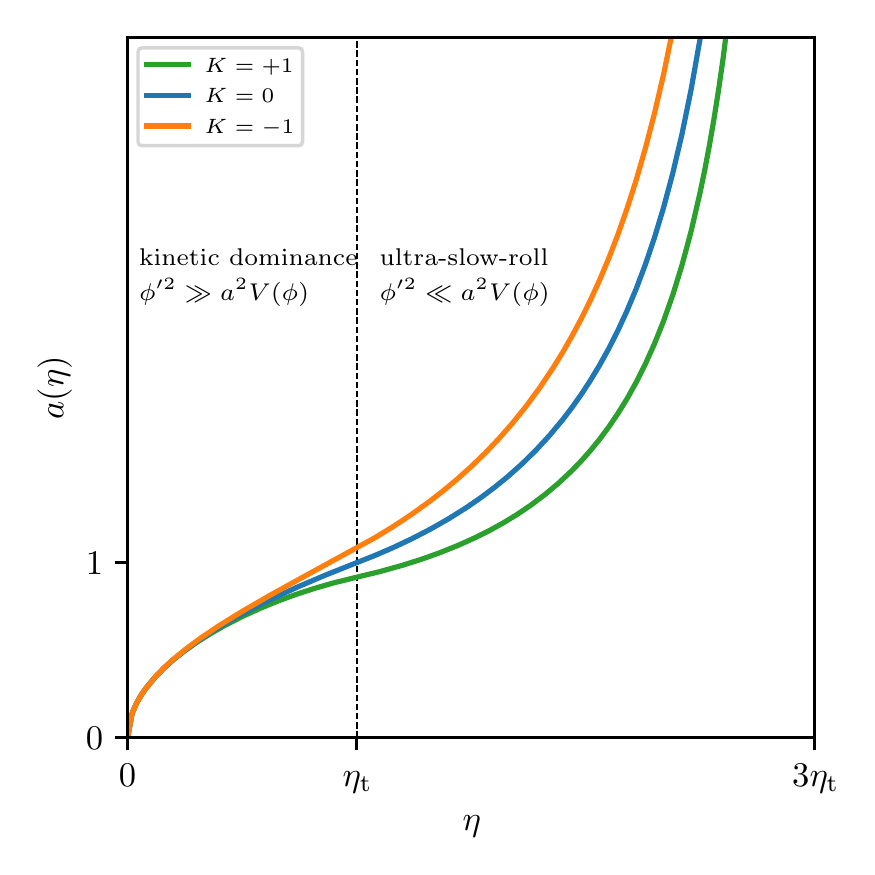}
    \caption{Evolution of the scale factor $a$ over conformal time from the analytical calculation in \cref{eqn:asol}, where the initial singularity has been set at $\eta = 0$ and the scale factor normalised  $a=1$ at the transition time $\etat$ for the case of a flat Universe ($K=0$).}
    \label{fig:scalefactor}
\end{figure}

If we define
\begin{equation}
    S_K(x) = \left\{
        \begin{array}{ll}
            \sin(x) & K=+1\\
            x & K=0\\
            \sinh(x) & K=-1,\\
        \end{array}
    \right.
    \label{eqn:sin}
\end{equation}
then solving \cref{eqn:kd} with the right-hand-side set to zero yields $a\sim \sqrt{S_K(2\eta)}$ for the kinetically dominanted regime, and solving \cref{eqn:usr} similarly yields $a\sim 1/S_K(\eta)$ for ultra-slow-roll. In both cases these solutions have two free integration constants corresponding to an additive coordinate shift in $\eta$ and a linear scaling of $a$. Matching $a$ and $a'$ for these two solutions at some transition time $\etat$ gives
\begin{equation}
    a = \left\{
        \begin{array}{ll}
            \sqrt{S_K(2\eta)} &:  0\le\eta<\etat\\
            {[S_K(2\etat)]^{3/2}}/{S_K(3\etat - \eta)} &: \etat\le\eta< 3\etat,\\
        \end{array}
    \right.
    \label{eqn:asol}
\end{equation}
with the conformal coordinate freezing out into the inflationary phase as $\eta \to 3\etat$. The evolution of the scale factor $a$ is plotted in \Cref{fig:scalefactor}.

Note that for the closed case ($K=+1$), there is a maximum sensible value of $\etat=\pi/4$. At values of $\etat$ greater than this, the Universe begins collapsing before the transition is reached and should be regarded as a breakdown of the approximation.

The remaining background variables may also be solved in the kinetically dominated regime with curvature and conformal time~\citep{kinetic_dominance}, but for the purposes of this analysis we only need power series expansions, which up to the first curvature terms read
\begin{align}
    N &= N_\p + \frac{1}{2}\log \eta - \frac{K}{3}\eta^2 + \mathcal{O}(\eta^4),\label{eqn:Nsol} \\
    \phi &= \phi_\p \pm \sqrt{\frac{3}{2}}\log \eta \pm \frac{\sqrt{6}K}{6}  \eta^2 + \mathcal{O}(\eta^4)\label{eqn:phisol},
\end{align}
where $N = \log a$. Other derived series include
\begin{align}
    \deriv{\phi} &=\pm\sqrt{\frac{3}{2}}\frac{1}{\eta} \pm \frac{\sqrt{6}K}{3}  \eta + \mathcal{O}(\eta^3),\label{eqn:phidotsol} \\
    \conformalH&= \deriv{N} = \frac{1}{2\eta} - \frac{2K}{3}\eta+ \mathcal{O}(\eta^3), \label{eqn:Hubblesol} \\
    a  &= e^N = e^{N_\mathrm{p}}\eta^{1/2}-\frac{e^{N_\mathrm{p}}K}{3}\eta^{5/2} + \mathcal{O}(\eta^{9/2}) \label{eqn:scalefactorsol}.
\end{align}

A complete derivation of these series requires a consideration of logolinear power series expansions~\citep{logolinear,lasenbyclosed}, which we detail further in \cref{sec:Appendix}.

The ultra-slow-roll regime is defined loosely as ${{\deriv{\phi}}^2 \ll a^2 V(\phi)}$, but can be more precisely thought of as the limit where $\mathcal{E}\to0$ but curvature contributions remain. For our analysis we only need the analytic form of the scale factor $a$ found in \cref{eqn:asol}.

\subsection{Mukhanov-Sasaki solutions under kinetic~dominance}\label{sec:Solving_CurvedMS_KD}

The evolution of the Mukhanov variable $v_k$ is defined by the Mukhanov-Sasaki~\cref{eqn:GeneralCurvedMS}. Combining the results from \cref{eqn:phidotsol,eqn:Hubblesol,eqn:scalefactorsol} show that for the kinetically dominated regime
\begin{equation}
    \frac{\dderiv{\mathcal{Z}}}{\mathcal{Z}} + 2 K + \frac{2K}{\conformalH}\frac{\deriv{\mathcal{Z}}}{\mathcal{Z}} = -\frac{1}{4\eta^2} + \frac{32K}{3} - \frac{24K^2}{\mathcal{K}^2(k)} + \mathcal{O}(\eta^{2}).
    \label{eqn:curvedMSKDfraction}
\end{equation}

Substituting (\ref{eqn:curvedMSKDfraction}) into (\ref{eqn:GeneralCurvedMS}) we can write the Mukhanov-Sasaki for the kinetically dominated regime as
\begin{gather}
    \dderiv{v}_k + \left[\kkd^2 + \frac{1}{4 \eta^2} \right] v_k = 0, \nonumber \\
    \kkd^2(k) = \mathcal{K}^2(k) - \frac{32 K}{3} + \frac{24 K^2}{\mathcal{K}^2(k)}.
    \label{eqn:curvedMSKDregime}
\end{gather}

From \cref{eqn:wavevectors,eqn:curvedMSKDregime} we can see that the first-order effects of curvature on the Mukhanov-Sasaki~\cref{eqn:GeneralCurvedMS} in the kinetically dominated regime manifest themselves purely as an effective shift in the wavevector participating dynamically.

By solving \cref{eqn:curvedMSKDregime} we find that during the kinetic dominance regime the Mukhanov variable $v_k$ evolves as
\begin{equation}
    v_k \left( \eta \right) = \sqrt{\frac{\pi}{4}} 
    \sqrt{\eta}  \left[A_k H_0^{(1)}(\kkd\eta) +B_k  H_0^{(2)}(\kkd\eta)\right],
    \label{eqn:generalKDsln} 
\end{equation}
where $H_0^{(1,2)}$ are zero-degree Hankel functions of the first and second kinds\footnote{and should not be confused with the present day Hubble constant}, and quantum mechanical normalisation requires $|B_k|^2-|A_k|^2=1$~\citep{quantum_initial_conditions}. Following \citet{Contaldi} and \citet{Sahni:1990tx} we choose initial conditions which select the right-handed mode
\begin{equation}
    A_k=0, \qquad B_k=1,\label{eqn:rhm}
\end{equation}
leaving a consideration of alternative quantum initial conditions to a future work.

\subsection{Mukhanov-Sasaki solutions under ultra-slow-roll}\label{sec:Solving_CurvedMS_USR}
For the ultra-slow-roll regime ($\eta\ge\etat$), taking the limit $\mathcal{E} \to 0$ shows that up to first order in curvature the relevant terms in the Mukhanov-Sasaki~\cref{eqn:GeneralCurvedMS} take the form
\begin{align}
    \frac{\dderiv{\mathcal{Z}}}{\mathcal{Z}}& + 2K + \frac{2K\deriv{\mathcal{Z}}}{\conformalH\mathcal{Z}} \to \frac{\dderiv{a}}{a} + 3K  \nonumber \\ 
    &= \frac{2}{(\eta - 3\etat)^2} + \frac{8K}{3} +  \mathcal{O}[(\eta - 3\etat)^2].
    \label{eqn:curvedMSUSRfraction}
\end{align}

Substituting this result from (\ref{eqn:curvedMSUSRfraction}) into (\ref{eqn:GeneralCurvedMS}), allows us to express the Mukhanov-Sasaki equation for the subsequent ultra-slow-roll regime as
\begin{gather}
    \dderiv{v}_k + \left[\kusr^2 - \frac{2}{(\eta - 3\etat)^2} \right] v_k = 0, \nonumber \\
    \kusr^2 = \mathcal{K}^2(k) - \frac{8K}{3}.
    \label{eqn:curvedMSUSRregime}
\end{gather}

Note that the shifted dynamical wavevector $k_+$ for ultra-slow-roll ($\eta\ge \eta_\mathrm{t})$ is distinct from that defined for the kinetically dominated regime $k_-$ ($\eta\le \eta_\mathrm{t}$).

By solving \cref{eqn:curvedMSUSRregime} we find that during the ultra-slow-roll stage, the Mukhanov variable $v_k$ evolves as
\begin{align}
    v_k(\eta) = \sqrt{\frac{\pi}{4}}\sqrt{3\etat-\eta}\Big[&C_k H^{(1)}_{{3}/{2}}(\kusr(3\etat - \eta)) \nonumber\\
    + &D_k H^{(2)}_{{3}/{2}}(\kusr(3\etat - \eta))  \Big].
    \label{eqn:generalLateSol}
\end{align}

One can now invoke the condition of continuity of $v_k$ and ${\deriv{v_k}}$ at the transition time $\etat$ and match \cref{eqn:generalKDsln,eqn:rhm,eqn:generalLateSol}, to show the coefficients of the two modes of the Mukhanov variable $v_k$ in the ultra-slow-roll regime, which are as follows:
\begin{align}
    C_k = \frac{i \pi\etat }{2\sqrt{2}} \Big[
      &\kusr H^{(2)}_{0}(\kkd \etat) \: H^{(2)}_{1/2}(2 \kusr \etat) \nonumber\\
    - &\kkd H^{(2)}_{1}(\kkd \etat) \: H^{(2)}_{3/2}(2 \kusr \etat)
\Big], \label{eqn:generalC}\\
    D_k = \frac{i\pi \etat}{2\sqrt{2}} \Big[
      &\kkd H^{(2)}_{1}(\kkd \etat) \: H^{(1)}_{3/2}( 2 \kusr \etat) \nonumber\\ 
    - &\kusr H^{(2)}_{0}( \kkd \etat) \: H^{(1)}_{1/2}( 2 \kusr \etat)
\Big]. \label{eqn:generalD}
\end{align}

This recovers the results obtained by \citet{Contaldi} in the limit of zero curvature ($K = 0$), \textit{i.e.} $\kkd^2 \to \mathcal{K}^2 \to k^2$ and $ \kusr^2 \to \mathcal{K}^2 \to k^2$. 

\subsection{The primordial power spectrum}\label{sec:PowerSpectrum}
With these complete solutions of the Mukhanov variable, we have the means to compute a primordial power spectrum.
By extending the analysis of~\citet{Contaldi} and generalizing to the curved case, we derive the curved primordial power spectrum of the comoving curvature perturbation $\mathcal{R}$ under our approximation to be~\footnote{\citet{Contaldi} considered the spectrum of a uniquely defined variable $Q$ in the limit where it becomes constant at late time.}

\begin{align}
    \mathcal{P}_\mathcal{R}(k) &\equiv \frac{k^3}{2 \pi^2} \vert \mathcal{R}_k \vert^2 \nonumber \\
    &\rightarrow \lim_{\eta\to3\etat}\frac{1}{8a^2 \mathcal{E}\pi^2(3\etat-\eta)^2}\frac{k^3}{\kusr^3} \vert C_k - D_k \vert^2, \nonumber \\ 
    &= A_s \frac{k^3}{\kusr^3} \vert C_k - D_k \vert^2,
    \label{eqn:PowerSpectrumR}
\end{align}
where we have used that $\mathcal{R}_k$ = $v_k/\mathcal{Z}_k$, and $\mathcal{Z}\to a\phi'/\conformalH = a\sqrt{2\mathcal{E}}$
where the transition time parameter $\etat$, slow-roll parameter $\mathcal{E}$ and formally diverging parameters can be absorbed into the usual scalar power spectrum amplitude $A_s$.
At short wavelengths, where $\kkd \to \kusr \to k \gg 1/\etat$, one recovers the standard result of a scale-invariant spectrum
\begin{equation}
    \vert C_k \vert \simeq 1,
    \qquad
    \vert D_k \vert \ll \vert C_k \vert, 
    \qquad
    \mathcal{P}_\mathcal{R} \simeq A_s.
    \label{largek}
\end{equation}

It should be noted that as we are working in the ultra-slow-roll regime, as in~\citet{Contaldi} there is no tilt $n_s$ to this power spectrum. Whilst there exist more sophisticated ways to incorporate higher order terms and hence recover the tilt, in this work we insert this by replacing $A_s$ with the standard tilted power spectrum parameterisation.

Our analytical form of the primordial power spectrum for each curvature $K\in\{+1,0,-1\}$ therefore is parameterised by an amplitude $A_s$, spectral index $n_s$ and transition time $\eta_\mathrm{t}$
\begin{equation}
\mathcal{P}_\mathcal{R}(k) = A_s {\left( \frac{k}{k_*} \right)}^{n_s-1} \frac{k^3}{\kusr^3} {\left\vert C_k(\eta_\mathrm{t}) - D_k(\eta_\mathrm{t}) \right\vert}^2,
    \label{eqn:PowerSpectrumR_final}
\end{equation}
where $C_k$ and $D_k$ are defined by \cref{eqn:generalC,eqn:generalD}, using Hankel functions and wavevectors $k_\pm$ defined in \cref{eqn:generalKDsln,eqn:curvedMSUSRregime}.

\begin{figure*}
    \centerline{\includegraphics{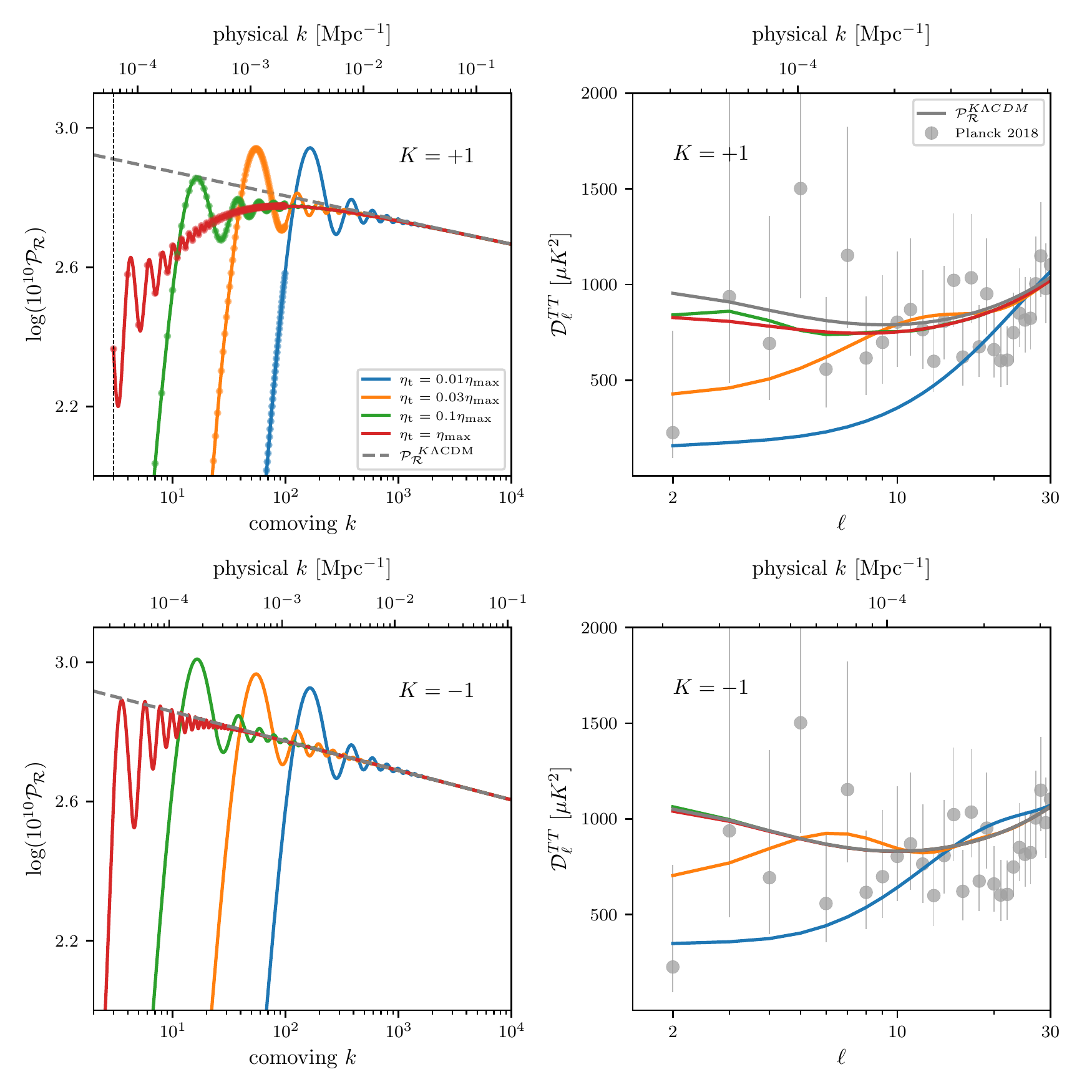}}
    \caption{Left: primordial power spectra $\mathcal{P}_\mathcal{R}$ corresponding to the range of allowed values of the transition time $\etat$ for open and closed universes $K\in\{-1,+1\}$. Oscillations and a generic suppression of power are visible at low-$k$. For $K = +1$, only integer values of comoving $k$ with $k \geq 3$ are allowed. Dots indicate the first $100$ comoving $k$. For clarity, we include the continuous spectrum. Right: the corresponding low-$\ell$ effects on the CMB power spectrum. The power law $K\Lambda$CDM spectrum is highlighted in grey along with Planck data. There is no appreciable deviation from the traditional power spectrum at higher $k$ and $\ell$ values. Note that the spectra of $\mathcal{P}_\mathcal{R}$ and $\mathcal{D}_\ell^{TT}$ qualitatively reproduce those found numerically in~\citep{Handley_2019}. Multipole $\ell$ and comoving \& physical $k$ are related by the conversions presented in \citet{Agocs_2020}.\vspace{30pt}}
    \label{fig:GeneralCurvedCMB2}
\end{figure*}

The spectra of $\mathcal{P}_\mathcal{R}$ generated by our analytical calculation are plotted in~\Cref{fig:GeneralCurvedCMB2}. We note that they reproduce the spectra obtained by~\citet{Contaldi} in the case of zero curvature ($K = 0$).

\section{Discussion}\label{sec:discussion}
Upon review of the calculations presented in \ref{sec:computingCurvedSpectra}, we see that when applying a purely analytical approach to solve curved inflationary dynamics, the effects of curvature can be mathematically attributed to shifts in the wavevectors participating dynamically \cref{eqn:curvedMSKDregime,eqn:curvedMSUSRregime}. Further inspection of the curved Mukhanov-Sasaki equation in \cref{eqn:conformalcurvedMS} provides a sanity check of this mathematical result, as we see that, within Fourier space, the differential operator is replaced by a scalar wavevector shifted by a curvature term. At a dynamic level we see that this shifted wavevector manifests itself in the spectra of $\mathcal{P}_\mathcal{R}$ as phase-based ringing effects for large enough values of the transition time, $\eta_{\mathrm{t}}$. This gives us a physical intuition of the oscillations seen in the numerically generated curved primordial power spectra for closed inflating universes~\citep{Handley_2019}. 

Furthermore, we have shown through our generally curved approach that curvature also manifests itself as a shifted wavevector in the open case, and thus these phase-based ringing effects are present in open inflating universes. Unlike the open case, for the closed case we do not obtain a sensible spectrum for all values of the transition time, $\eta_t$; for large $\eta_t$, we observe a natural breakdown of the approximation at low $k$ at the limit of $k=3$ for comoving $k$. This is in agreement with the constraint $k \in \Z>2$ for closed universes, below which the frequency of the oscillatory solutions become imaginary. In the spectrum of~\citet{Contaldi} generated for flat $\Lambda$CDM, there is also a ringing effect sourced by the instantaneous transition which causes a discontinuity in the Mukhanov-Sasaki fraction $\dderiv{z}/z$. The flat ($K=0$) spectra generated by our general curved approach also demonstrate these effects, but in the curved case there is a discontinuity in $\dderiv{\mathcal{Z}}/\mathcal{Z} + 2K + 2K\deriv{Z}/\conformalH\mathcal{Z}$, which in the flat case ($K=0$) reduces to a discontinuity of $\dderiv{z}/z$.

$K\Lambda$CDM is a commonly considered extension to standard flat $\Lambda$CDM, where there is an additional degree of freedom of spatial curvature $\Omega_K$. In $K\Lambda$CDM an almost flat power spectrum is assumed
\begin{equation}
    \mathcal{P}_\mathcal{R}^{K\Lambda\mathrm{CDM}}(k) = A_s \left( \frac{k}{k_*} \right)^{n_s-1},
    \label{eqn:PPS_Klcdm}
\end{equation}
where $k_*$ corresponds to the pivot perturbation mode and by convention is set to have a length-scale today of $0.05 \text{Mpc}^{-1}$. 

The \Planck{} 2018 results including CMB lensing give the curved universes (TTTEEE+lowl+lowE+lensing) best-fit data as $A_s = 2.0771\pm0.1017 \times 10^{-9}$ and $n_s = 0.9699 \pm 0.0090$. Hence, the observations support a weak power law decay of the primordial power spectrum.

\citet{Handley_2019} showed that relative to \cref{eqn:PPS_Klcdm}, including the exact numerical calculation for curved universes, introduces oscillations and a suppression of power at low $k$, independent of initial conditions, and hence deviates from the form of \cref{eqn:PPS_Klcdm}. It has been shown in previous work that the $(k/k_*)^{n_s-1}$ tilt is a higher-order effect manifested from the nature of the scalar field potential chosen for the slow-roll regime. To compute such effects, one can determine the higher order terms of the {\em logolinear expansions} listed in \cref{sec:Appendix}.

As a good phenomenological approximation for a general inflationary setting, we scale our normalised $\mathcal{P}_\mathcal{R}$ with the best-fit scalar power spectrum amplitude $A_s$ and manually add in the tilt, we present our analytical primordial power spectrum $P_\mathcal{R}$ for varying values of the transition time $\etat$, which we then follow through to the CMB, in~\Cref{fig:GeneralCurvedCMB2}~\cite{CLASS2}. The CMB spectra, corresponding to these primordial power spectra, are generated using parameters values set in accordance with the best-fit data for each curved scenario. For the closed case we use the \Planck{} 2018 TTTEEE+lowl+lowE+lensing best-fit parameters. For the flat case we work with the best-fit parameters for a flat $\Lambda$CDM cosmology. For the open case we calculate the mean posterior distribution of all lensing data using the \texttt{anesthetic} package, subject to the constraint that $\Omega_k > 0$ ($K=-1$)~\cite{anesthetic}.

The requirement that the horizon problem is solved \textit{i.e.} that the amount of conformal time during inflation $\eta_{\mathrm{i}}$ is greater than the amount of conformal time before $\etat$ and afterward $\eta_\mathrm{\uparrow}$, bounds the transition time $\etat$ from above. The condition of the amount of conformal time during inflation $\eta_{\mathrm{i}}$ being greater than the conformal time in the kinetically dominated regime preceding inflation $\etat$ is naturally satisfied by the ultra-slow-roll solution of (\ref{eqn:asol}), since $\eta_{\mathrm{i}}=2\eta_{\mathrm{t}}$. The additional condition regarding the conformal time after inflation $\eta_\mathrm{\uparrow}$ places the constraint that ($\eta_\mathrm{\uparrow}<2\etat$). The implications of these constraints on exact numerical integration methods for computing curved primordial power spectra, are discussed in more detail in~\citet[Section IV]{Handley_2019}.

Figure \ref{fig:GeneralCurvedCMB2} demonstrates how the computed spectra vary for different values of the transition time $\etat$. The location of the cutoff, suppression of power and oscillations are changed by adjusting the transition time, and as expected~\cite{lasenbyclosed} the depth of the suppression in closed universes ($K=+1$) is greater for the case when primordial curvature has a larger magnitude (corresponding to a higher transition time). Interestingly, we find that for large enough values of the transition time, a suppression of power is also seen in open universes ($K=-1$). 

Overall, we demonstrate that our analytical calculations reproduce very well the spectra obtained with the exact numerical evolution reported in~\citep{Handley_2019}, as well as the spectrum obtained by the analytical approximation of \citet{Contaldi}, \textit{i.e.} the case of zero curvature ($ K = 0 $). With this work we have not only developed an analytical framework to solve curved inflationary dynamics, but a means to study curvature in isolation, without complicating factors, such as the choice of the scalar field potential.

\section{Conclusions}\label{sec:conclusions}
The inflationary scenario addresses the initial value problem of the Hot Big Bang, but provides us with no insight into the Universe's state pre-inflation. Therefore, in order to truly understand the physics of inflation, we must study it with no bias toward the conditions of the Universe at inflation start; more specifically, we can not infer the shape of the Universe prior to inflation from the observed flatness seen at recombination through the CMB.

In~\citep{Handley_2019}, it was shown through exact numerical calculations of curved inflating universes generated spectra with generic cut-offs and oscillations within the observable window for the level of curvature allowed by current CMB measurements and provide a better fit to current data. In this work we have used the formalism popularised by~\citep{1992PhR...215..203M, Lesgourgues:2013bra, Baumann} and subsequent manipulation to write the Mukhanov-Sasaki equation for curved universes in conformal time. This has allowed us to derive analytical solutions of the Mukhanov-Sasaki equation for a generally curved Universe scenario, which show that curvature mathematically manifests itself as a shifted dynamical wavevector, and physically at low $k$ as a suppression of power and oscillations in the primordial power spectra, which then follow through to the CMB.

The main emphasis of our paper was related to modifications of the simple model utilised by \citet{Contaldi}, which invokes an instantaneous transition between an initial kinetic stage (when the velocity of the scalar field was not negligible) and an approximate de Sitter inflationary stage, to generate the significant suppression of the large scale density perturbations. Through the application of logolinear series expansions and a newly defined inflationary regime, we generalise this model to the curved inflating case, to introduce oscillations and a suppression of power at low $k$, as well as generic cut-offs, which is in agreement with exact numerical calculations. Varying the remaining degree of freedom, specifically the amount of primordial curvature (provided through the transition time), alters the oscillations and level of suppression in a non-monotonic manner, whilst there is a consistent lowering in the position of the cut-off at low $k$ with increasing transition time.

The addition of an extra curvature parameter in the theory to obtain a better fit with data comes with costs, but given the recent discrepancies that have arisen with the standard $\Lambda$CDM model, this is something that now requires strong consideration. A natural extension is $K\Lambda$CDM. Our work has now shown, both analytically and numerically, that for all allowed values of initial primordial curvature, incorporating the exact solutions for closed universes results in observationally significant alterations to the power spectrum. Furthermore, the data are capable of distinguishing a preferred vacuum state, with the best fit preferring renormalised-stress-energy-tensor (RSET) initial conditions over the traditional Bunch--Davies vacuum. Future work will involve extending our analytical approach to RSET and other initial conditions.

\begin{acknowledgments}
    A.T. dedicates this paper to the bright memory of Rafael Baptista Ochoa. D.W. thanks the ENS Paris-Saclay for its continuing support via the normalien civil servant grant. W.H.\ would like to thank Gonville~\&~Caius College for their ongoing support. The authors would like to thank Fruzsina Agocs, Julien Lesgourges and Thomas Gessey-Jones for their conversations on the nature of curved primordial power spectra. Josh Da Cruz, Megan Pritchard, Nicholas J. Cooper, Oliver Philcox and two anonymous reviewers provided helpful comments on earlier drafts of the paper.
\end{acknowledgments}

\appendix
\section{Logolinear expansions in conformal time}\label{sec:Appendix}
Logolinear series expansions~\citep{logolinear} for a general function $x(\eta)$ have the form
\begin{equation}
    x(\eta) = \sum_{j,k} [x^k_j] \: \eta^j {\left( \log \eta \right)}^k,
    \label{eqn:logolineardefinition}
\end{equation}
where $[x^k_j]$ are twice-indexed real constants defining the series, with square brackets used to disambiguate powers from superscripts.

We begin with \cref{eqn:raychaudhuri,eqn:klein_gordon}
\begin{align}
    \dderiv{N} + {\deriv{N}}^2 +\frac{1}{3}\left( {\deriv{\phi}}^2 - a^2V(\phi) \right) &=0,
    \label{eqn:raychaudhuri_N}\\
    \dderiv{\phi} + 2 \deriv{N}\deriv{\phi} + a^2\frac{\d{}}{\d{\phi}}V(\phi) &=0.
    \label{eqn:klein_gordon_N}
\end{align}

Here $N = \log a$ has been used rather than $\conformalH$ as it restates \cref{eqn:raychaudhuri,eqn:klein_gordon} in the form of second order differential equations, which we can then in turn convert to a first order system of equations
\begin{align}
    \dot{N} &= h,
    \qquad 
    \dot{\phi} = v,
    \nonumber\\
    \dot{h} &= h -\frac{1}{3} v^2 + a^2\frac{1}{3} \eta^2V(\phi),
    \nonumber\\
    \dot{v} &= v - 2 v h - a^2\eta^2\frac{\d{}}{\d{\phi}}V(\phi),
    \label{eqn:dsys}
\end{align}
where dots indicate derivatives with respect to logarithmic conformal time $\log\eta$, \textit{i.e.} $\dot{x}=\frac{\d{}}{\d{\log \eta}} x$.

To analytically determine approximate solutions for curved cosmologies we will consider series expansions for a general function $x(\eta)$ of the form \footnote{Note that this indexing convention differs from that adopted in~\citep{lasenbyclosed}, which also utilised series expansions to solve cosmological evolution equations. For our purposes an expansion in $\eta$ was required, hence the unique convention used in our series definitions.}
\begin{equation}
    x(\eta) = \sum_j x_j(\eta)\: \eta^j \quad\Rightarrow\quad \dot{x}(\eta) = \sum_j (\dot{x}_j + j x_j)\: \eta^j.
    \label{eqn:logolinear}
\end{equation}

Substituting in our series definition from \cref{eqn:logolinear} and equating coefficients of $\eta^j$, we find that \cref{eqn:dsys} becomes
\begin{align}
    \dot{N}_j + j N_j &= h_j,
    \qquad
    \dot{\phi}_j + j \phi_j = v_j,
    \nonumber\\
    \dot{h}_j + j h_j &= h_j + \frac{1}{3} V(\phi)e^{2N_\p}e^{\sum_{q>0}N_q(\eta)\eta^q}|_{j-3} -\smashoperator{\sum_{p+q=j}}\frac{v_p v_q}{3},
    \nonumber\\
    \dot{v}_j + j v_j &= v_j - \frac{\d{V(\phi)}}{\d{\phi}}e^{2N_\p}e^{\sum_{q>0}N_q(\eta)\eta^q}|_{j-3} - 2 \smashoperator{\sum_{p+q=j}} v_p h_q.
    \label{eqn:dsysj}
\end{align}

One should also consider the equivalent of \cref{eqn:friedmann}
\begin{equation}
    \frac{1}{3}V(\phi)e^{2N_\p}e^{\sum_{q>0}N_q(\eta)\eta^q}|_{j-3} +\smashoperator{\sum_{p+q=j}}{\frac{1}{6}}v_p v_q - h_p h_q =K|_{j-2},
    \label{eqn:friedmann_logt}
\end{equation}
where exponentiation of logolinear series was discussed in~\citep{logolinear}.

We may solve for the $j=0$ case of \cref{eqn:dsysj} using the kinetically dominated solutions, as it is equivalent to ~\cref{eqn:dsys} with $V=0$
\begin{align}
    N_0 &= N_\p + \frac{1}{2}\log \eta, &h_0 &= \frac{1}{2}, \nonumber\\
    \phi_0 &= \phi_\p \pm \sqrt{\frac{3}{2}}\log \eta, &v_0 &=\pm\sqrt{\frac{3}{2}},
    \label{eqn:0_sol}
\end{align}
where $N_p$ and $\phi_p$ are constants of integration. As mentioned previously we expect there to be four constants of integration {\em a priori}. One of the missing constants is fixed by defining the singularity to be at $\eta=0$, whilst the other is effectively set by the curvature. Hence \cref{eqn:0_sol} represents a complete solution to $j=0$ for only the flat case ($K=0$). Nevertheless, we may still adopt \cref{eqn:0_sol} as the base term for the logolinear series. The final constant of integration will then effectively emerge from a consideration of higher order terms.

For $j\ne 0$, we can rewrite \cref{eqn:dsysj} in the form of a first order linear inhomogeneous vector differential equation
\begin{equation}
    \dot{x}_j + A_j x_j = F_j,
    \label{eqn:linear_master}
\end{equation}
where $x=(N,\phi,h,v)$, $A_j$ is a (constant) matrix
\begin{align}
    A_j &= \left(
    \begin{array}{cccc}
        j & 0 & -1 & 0 \\
        0 & j & 0 & -1 \\
        0 & 0 & j-1 & \frac{2}{3}v_0 \\
        0 & 0 & 2v_0 & j-1+2h_0 \\
    \end{array}
    \right),\nonumber\\
    &= \left(%
    \begin{array}{cccc}
        j & 0 & -1 & 0 \\
        0 & j & 0 & -1 \\
        0 & 0 & j-1 & \pm\sqrt{\frac{2}{3}} \\
        0 & 0 & \pm\sqrt{6} & j \\
    \end{array}
    \right),\label{eqn:A}
\end{align}
and $F_j$ is a vector polynomial in $\log \eta$ depending only on earlier series $x_{p<j}$
\begin{align}
    F_j &=
    \left(
    \begin{array}{c}
        0\\
        0\\
        \frac{1}{3} V(\phi)e^{2N_\p}e^{\sum_{q>0}N_q(\eta)\eta^q}|_{j-3}-\smashoperator{\sum_{\substack{p+q=j\\p\ne j,q\ne j}}} \frac{1}{3}v_p v_q \\
        - \frac{\d{V(\phi)}}{\d{\phi}}e^{2N_\p}e^{\sum_{q>0}N_q(\eta)\eta^q}|_{j-3} - 2 \smashoperator{\sum_{\substack{p+q=j\\p\ne j,q\ne j}}} v_p h_q 
    \end{array}
    \right).\label{eqn:Fj}
\end{align}

At each $j$, the linear differential \cref{eqn:linear_master} may be solved in terms of a complementary function ($x_j^\mathrm{cf}$) with four free parameters and a particular integral ($x_j^\mathrm{pi}$), \textit{i.e.} $x_j = x_j^\mathrm{cf} + x_j^\mathrm{pi}$. These free parameters correspond to the degrees of gauge freedom mentioned in~\citep{logolinear}.

We may solve the homogeneous version of \cref{eqn:linear_master} exactly, since $A_j$ is a constant matrix
\begin{equation}
    \frac{\d{x_j^\mathrm{cf}}}{\d{\log \eta}} + A_j x_j^\mathrm{cf} = 0  \quad\Rightarrow\quad x_j^\mathrm{cf} = e^{-A_j\log \eta}[x_j^0],
\end{equation}
where $[x_j^0]$ is a constant vector parametrising initial conditions.
To compute the matrix exponential, we first compute eigenvectors and eigenvalues of $A_j$
\begin{align}
    e_\sft &= \left(%
    \begin{array}{cccc}
        1& \pm\sqrt{6} & \frac{(\sqrt{6}-18)}{12}& \mp\sqrt{6} \\
    \end{array}
    \right),& 
    A_j e_\sft{} &= (j+1)\cdot e_\sft,
    \nonumber\\
    e_b &= \left(%
    \begin{array}{cccc}
        1& \mp\frac{\sqrt{6}}{2} & \frac{(\sqrt{6}+18)}{12}& \mp\sqrt{6} \\
    \end{array}
    \right),&
    A_j e_b &= (j-2)\cdot e_b,
    \nonumber\\
    e_n &= \left(%
    \begin{array}{cccc}
        1& 0& 0& 0\\
    \end{array}
    \right),&
    A_j e_n &= j\cdot e_n,
    \nonumber\\
    e_\phi{} &= \left(%
    \begin{array}{cccc}
        0& 1& 0& 0\\
    \end{array}
    \right),&
    A_j e_\phi{} &= j\cdot e_\phi.\label{eqn:eigenvalues}
\end{align}

Parametrising initial conditions $[x_j^0]$ using the eigenbasis in \cref{eqn:eigenvalues} with parameters $\tilde{N},\tilde{\phi},\tilde{b}, \tilde{\sft}$, yields
\begin{align}
    x_j^\mathrm{cf}
    &= e^{-A_j\log \eta}(\tilde{N}e_n + \tilde{\phi}e_\phi + \tilde{b} e_b + \tilde{\sft} e_\sft)\nonumber\\
    &= \left(\tilde{N}e_n + \tilde{\phi}e_\phi + \tilde{b}e_b \eta^{2} + \tilde{\sft} e_\sft \eta^{-1}\right)\eta^{-j}.
    \label{eqn:complementary_function}
\end{align}

We may absorb all $\tilde{N}$ and $\tilde{\phi}$ into our definitions of $N_\p$ and $\phi_\p$. Choosing $\tilde{\sft}=0$ amounts to setting the singularity to be at $\eta=0$ as an initial condition without loss of generality, as it grows faster than our leading term as $\eta\to0$. The only remaining undetermined integration constant is $\tilde{b}$, which amounts to the integration constant that was missing from \cref{eqn:0_sol}. The constant $\tilde{b}$ is controlled by the curvature of the Universe via \cref{eqn:friedmann_logt}
\begin{equation}
    \tilde{b} = -\tfrac{1}{3}K .
    \label{eqn:curvature_relation}
\end{equation}

Applying the standard definition of conformal time $\mathrm{d}\eta=\mathrm{d}t/a$, shows a clear equivalence between \cref{eqn:curvature_relation} and the cosmic time version found in the series solutions derived in~\citep{logolinear}. We can now exchange $K$ for $\tilde{b}$ via this relation, and for the proceeding analysis in the main body of the paper we shall drop the notation of $\tilde{b}$ and explicitly denote curvature terms with $K$ in the series solutions. We also note from (\ref{eqn:complementary_function}) that the curvature of the Universe depends on a term in $\eta^2$.

All that remains to be determined is a particular integral of \cref{eqn:linear_master}, given that one has the form of $F_j$ at each stage of recursion.
The trial solution is $x_j(\eta)=\sum_{k=0}^{N_j} [x^k_j] {(\log \eta)}^k$. Defining ${F_j =\sum_{k=0}^{N_j} [F^k_j] {(\log t)}^k}$ and equating coefficients of ${(\log \eta)}^k$ gives
\begin{equation}
    (k+1)[x^{k+1}_j] + A_j [x^k_j] = [F^k_j],
    \label{eqn:linear_master_no_j}
\end{equation}
giving a descending recursion relation in $k$
\begin{equation}
    [x^{N_j+1}_k]=0,\quad [x^{k-1}_j] = A_j^{-1}( [F_j^{k-1}]  - k [x^{k}_j]).
    \label{eqn:recursion_relation}
\end{equation}

The recursion relation in \cref{eqn:recursion_relation} fails when $A_j$ is non-invertible, which occurs when any of the eigenvalues in \cref{eqn:eigenvalues} are zero ($j=-1,0,2$). For these cases, the system is underdetermined, with an infinity of solutions parameterised along the directions of relevant eigenvectors. This infinity of solutions can therefore be carefully absorbed into a corresponding constant of integration.

Similarly, if we were to define an alternative base to the recursion in \cref{eqn:recursion_relation}, then infinite series would be generated. However, all but a finite number of terms would merely contribute to a redefinition of constants $N_\p$, $\phi_\p$, $\tilde{b}$, or an introduction of non-zero $\tilde{\sft}$, which we disallow due to the consequent shift of the singularity to a non-zero conformal time $\eta$.

\bibliographystyle{unsrtnat}
\bibliography{references}

\end{document}